\documentclass[fleqn,usenatbib]{mnras}

\usepackage{newtxtext,newtxmath}

\usepackage[T1]{fontenc}

\usepackage{scalerel}
\usepackage{tikz}
\usetikzlibrary{svg.path}

\DeclareRobustCommand{\VAN}[3]{#2}
\let\VANthebibliography\thebibliography
\def\thebibliography{\DeclareRobustCommand{\VAN}[3]{##3}\VANthebibliography}

\usepackage{graphicx}	
\usepackage{amsmath}	

\definecolor{orcidlogocol}{HTML}{A6CE39}
\tikzset{
  orcidlogo/.pic={
    \fill[orcidlogocol] svg{M256,128c0,70.7-57.3,128-128,128C57.3,256,0,198.7,0,128C0,57.3,57.3,0,128,0C198.7,0,256,57.3,256,128z};
    \fill[white] svg{M86.3,186.2H70.9V79.1h15.4v48.4V186.2z}
                 svg{M108.9,79.1h41.6c39.6,0,57,28.3,57,53.6c0,27.5-21.5,53.6-56.8,53.6h-41.8V79.1z M124.3,172.4h24.5c34.9,0,42.9-26.5,42.9-39.7c0-21.5-13.7-39.7-43.7-39.7h-23.7V172.4z}
                 svg{M88.7,56.8c0,5.5-4.5,10.1-10.1,10.1c-5.6,0-10.1-4.6-10.1-10.1c0-5.6,4.5-10.1,10.1-10.1C84.2,46.7,88.7,51.3,88.7,56.8z};
  }
}

\newcommand\orcid[1]{\href{https://orcid.org/#1}{\mbox{\scalerel*{
\begin{tikzpicture}[yscale=-1,transform shape]
\pic{orcidlogo};
\end{tikzpicture}
}{|}}}}

\title[Lensing cosmic drift]{Lensing cosmic drift}

\author[G. Covone \& M. Sereno]{
Giovanni Covone
\orcid{0000-0002-2553-096X}
$^{1,2,3}$\thanks{E-mail: giovanni.covone@unina.it (GC)}
and Mauro Sereno \orcid{0000-0003-0302-0325}$^{4,5}$\thanks{E-mail: mauro.sereno@inaf.it (MS)}
\\
$^{1}$Dipartimento di Fisica "Ettore Pancini", Università di Napoli Federico II, Napoli, Italy\\
$^{2}$INFN, Sezione di Napoli, Complesso Universitario di Monte S. Angelo,
Via Cintia Edificio 6, 80126 Napoli, Italy \\
$^{3}$INAF - Osservatorio Astronomico di Capodimonte,
via Moiariello 16, 80131 Napoli, Italy \\
$^{4}$INAF -- Osservatorio di Astrofisica e Scienza dello Spazio di Bologna, via Piero Gobetti 93/3, I-40129 Bologna, Italy \\
$^{5}$INFN, Sezione di Bologna, viale Berti Pichat 6/2, 40127 Bologna, Italy\\
}

\pubyear{2022}

\begin{document}
\label{firstpage}
\pagerange{\pageref{firstpage}--\pageref{lastpage}}
\maketitle

\begin{abstract}
As the Universe expands, the redshift of distant sources changes with time. Here we discuss gravitational lensing phenomena that are consequence of the redshift drift between lensed source, gravitational lens, and observer. When the source is located very close to the drifting caustics, a pair of images could occur (or disappear) because of the cosmological expansion. Furthermore, lensing systems act as signal converters of the redshift drift. The angular position, magnification, distortion, and time delay of already existing multiple images change. We estimate the expected frequency of these phenomena and the prospects to observe them in the era of deep and large surveys. The drift detection in image separation could be within reach of next generation surveys with $\mu$arcsec angular resolution.
\end{abstract}




\section{Introduction}

The redshift drift is the time variation of the cosmological redshift of distant sources due to the expansion of the Universe. This subtle physical effect was predicted by Sandage \citep{san62} and in an homogeneous cosmological model it is described by the  equation
\begin{equation}
\label{eq_der_z}
\frac{d z }{d t_0}=H_{0}(1+z)-H(z) \, ,
\end{equation}
where $t_0$ is the time at the observer, $H_0$ is the  value of the Hubble-Le\^{i}matre constant today, and $H(z)$ is the value of the Hubble-Le\^{i}matre constant at the redshift $z$ of the source.

The  spectral shift is of the order of $10^{-10}$ over a decade, but  the observational confirmation appears to be within reach of next generation high-resolution spectrographs to be installed on the Extremely Large Telescope and other 30m-class telescopes, see, e.g., \citet{pas+al05}.
These experiments will target bright high-redshift quasars and aim at measuring the redshift variations in the complex systems of absorption lines due to neutral hydrogen along the line of sight \citep{loe98}.

The successful observation of the redshift drift would be of paramount importance as it would represent  the first direct evidence of the expansion of the Universe and a direct measurement of the expansion rate. 

Here we present a novel approach to observe the redshift drift by taking advantage of gravitational lensing. Strong gravitational lenses (as massive galaxies or galaxy clusters) produce magnified and distorted multiple images of background sources, sometimes visible as giant bright arcs (i.e., gravitational arcs). The observed properties of the lensed images at high magnifications $\mu$ (i.e., $\mu$ larger than a few tens) depend critically on the caustics structure and the angular separation of the source to the caustics. This implies that even tiny variations of the source position with respect to the caustics could produce observable effects. Redshift drift makes distances change and hence impact the magnification pattern.

The order of magnitude of any effect related to the redshift drift is determined by the quantity $\delta t_0 H_0$, where $\delta t_0$ is the time span of the experiment carried out by the observer. For $\delta t_0 \sim 10 \, {\rm yr}$, that is the expected duration of a typical observational campaign, $\delta t_0  H_0 \sim 7 \times 10^{-10}$. This makes measurements of cosmic drift very challenging.

\citet{loe98} suggested to measure the frequency or redshift shift induced by the Hubble flow between multiple images of lensed sources. \citet{pi+gi17} proposed to measure the positional shift of multiple images of lensed sources, or the change in time delay between them, over a certain period. 

These ideas were initially regarded as impractical given the typical spectroscopic sensitivity but some lensing systems might make the measurements feasible. \citet{zi+ei18} considered strongly lensed fast radio bursts (FRBs) to observe cosmological evolution in real time. The duration of a FRB is very short ($\sim 10^{-3}~\text{s}$), and the change in time delay between a pair of multiple image could be a measurable effect. Strongly lensed repeating FRBs were also considered as probes of dark energy evolution \citep{liu+al19}.

\citet{wuc+al21} further discussed time delay changes over time as a consequence of the variations of the redshift of both the gravitational lens and the background lensed source and the possibility of observing them on a feasible time scale. They argued that coherent time-delay measurements of gravitationally lensed FRBs can be measured to levels of accuracy that are better than the burst duration. This would allow us to disentangle cosmological effects from the proper motion and to eliminate the unknown mass distributions of the lensing galaxies as the main source of potential systematic errors.

In this paper, we revisit some of these ideas considering more flexible lens models, propose some novel observable, and discuss the prospects of observing them in future wide-area surveys. 

We adopt a flat $\Lambda$CDM cosmological model, with the following values of the cosmological parameters: $\Omega_\text{M} = 0.3$, $\Omega_{\Lambda} = 0.7$, and $H_0 = 70 \, {\rm km/s} \, {\rm Mpc}^{-1} \, .$
%
However, the results presented here do not depend critically on this choice.


\begin{figure}
\centering
\includegraphics[width=\columnwidth]{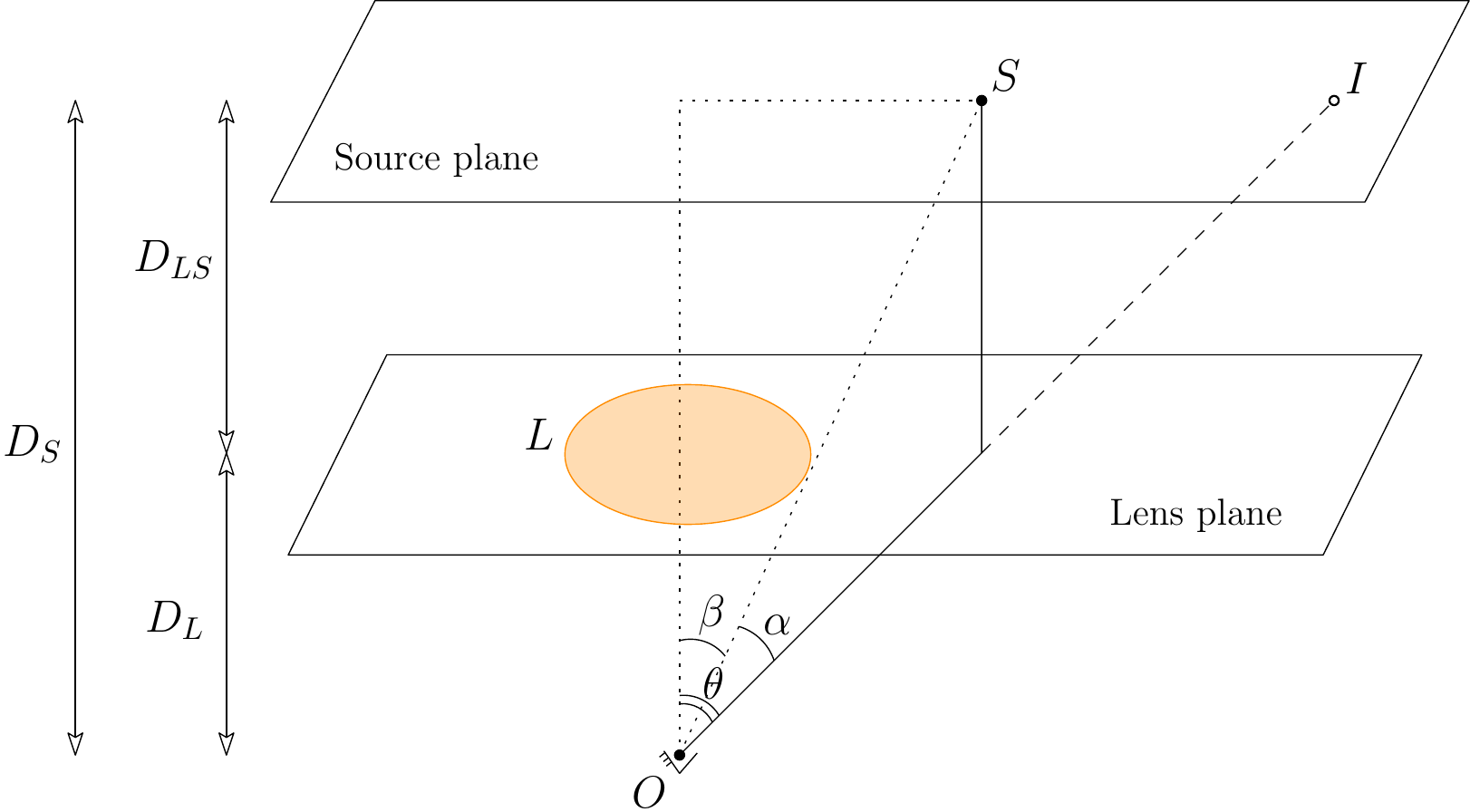}
\caption{Configuration of a simple gravitational lensing system, consisting of an axially symmetric gravitational lens $L$ and a source $S$ at cosmological distances from the observer $O$.}
\label{fig:one}
\end{figure}


\section{A simple model}

Let us consider an axially symmetric gravitational lens, located at redshift $z_\text{l}$, magnifying a background source at redshift $z_\text{s}$, as shown in Fig. \ref{fig:one}. We adopt the formalism introduced by \citet{sef92}. 
The vector angles $\bmath{\beta}$ and $\bmath{\theta}$ indicate the unlensed source and the image position, respectively, on their own planes. Angles are measured from the axis connecting the observer and the centre of the gravitational lens.


The quantities $D_\text{l}$, $D_\text{s}$, and $D_\text{ls}$ are the angular diameter distances of the gravitational lens, the source, and between the lens and the source, respectively. In a given cosmological model, the angular-diameter distances are known function of the redshift and the cosmological parameters.

To our knowledge, how cosmic drift impacts lensing phenomena is still a relatively novel topic. As a first step, we consider the setting of a thought experiment. We neglect the proper motions of the observer, the gravitational lens, and the background sources. That is, we consider them to be comoving objects, at rest in their local reference frame, see for instance \cite{har00}. 

We focus on the effects of redshift drift and we assume that the lens structure does not change with time. Lensing phenomena are mostly determined by the mass density within the scale radius of galaxies and clusters, which remains approximately constant during the slow accretion stage which follows an early fast accretion phase and are decoupled from the cosmic expansion \citep{di+jo19}. 
For our main analysis, we neglect mergers or accretion processes in the outskirts, as well as orbits and internal motion of internal components.

The expansion of the Universe causes a change in the global geometry of the lensing system, as the angular-diameter distances vary as a consequences of the redshift change. It is important to note that in an isotropic universe the source position angle $\bmath{\beta}$ does not change in time: indeed, it coincides with the comoving angular coordinate used in the Robertson-Walker metric. 
 

The configuration in Fig. \ref{fig:one} represents the lensing geometry at the observer cosmic time $t_0$. 
In the following, we consider the configuration of the same system at a later cosmic time $t_0 + \delta t_0$, when the distances are changed because of a variation in the scale factors, as a consequence of the cosmic expansion. We remind that the variation of the angular diameter distance $D_\text{A}$ due to drift can be written as \citep{zi+ei18,wuc+al21}
\begin{equation}
\frac{\text{d} \ln D_\text{A}(z)}{\text{d} t_\text{0}} = \frac{H(z)}{1+z} \, .
\label{eq_der_dist}
\end{equation}
From Eqs.~(\ref{eq_der_z},~\ref{eq_der_dist}), it follows that
\begin{equation}
\frac{\text{d}}{\text{d} t_\text{0}}\frac{D_\text{ls}}{D_\text{s}} = 0 \, ,
\end{equation}
and
\begin{equation}
\frac{\text{d}}{\text{d} t_\text{0}} \left[ (1+z )D_\text{A}\right] = H_0 \, .
\end{equation}
Following \citet{zi+ei18,wuc+al21}, we do not normalise the present-day scale factor $a_0$ to unity, because in real-time cosmology we need to consider that $a_0$ changes with time. As noted in \cite{zi+ei18}, this may explain some disagreement with previously published results.

The so-called lensing equation is straightforwardly derived from the configuration in Fig. \ref{fig:one}:
\begin{equation}
\bmath{\beta} \, = \bmath{\theta} - \bmath{\alpha} (\bmath{\theta}) \, . 
\label{eq:lensing-equation}
\end{equation}
This equation defines a 2D mapping between the image plane and the source plane. Caustics and critical lines (in the source  and  image  planes,  respectively) are the places where the Jacobian of the map vanishes, see for instance \citet{sef92}. 
The variation of the lensing geometry causes a variation in the shape of caustics and critical lines and then could manifest via several observational effects. 
Here we only consider smooth mass models, i.e. we do not consider the effect of the expanding caustics on the microimages produced by the
not-smooth distribution within the gravitational lenses.

In particular, we focus on the following observational features: 
(a) the appearance (or disappearance) of new images for any source which gets included (or ousted) by the caustics as the enclosed area gets larger (or smaller),
(b) the variation in time of the image separation, 
(c) the change in the time-delay between the images, 
and, finally, (d) the effect on shear measurements in the weak lensing regime.


\begin{figure}
\centering
\includegraphics[width=\columnwidth]{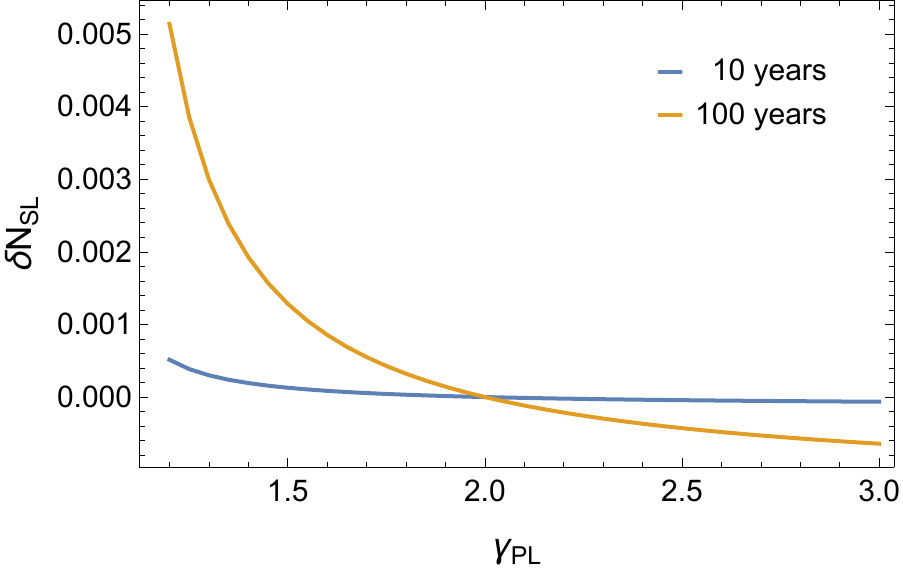}
\caption{Number $\delta N_\text{SL}$ of newly occurred images due to cosmic drift in a survey with $10^5$ self-similar lenses at $z_\text{l}=0.3$ as a function of the power slope index $\gamma_\text{PL}$ over a span of 10 (cyan) or 100 (orange) years.}
\label{fig_delta_NSL_gamma_PL}
\end{figure}


\subsection{Occurrence of new images}

A new pair of images form when the background source moves inward a caustic. As the universe expands and the redshifts of the source and gravitational lens change, the cosmological distances involved in the geometry of the lensing system change as well. As a consequence the caustics and the critical lines are slightly modified. For an axially symmetric gravitational lens, we expect an isotropic variation of the critical lines as the Einstein radius is a function of  $z_\text{l}$ and $z_\text{s}$.

Here, we assume that the gravitationally bound lens is detached from the cosmic expansion and its matter distribution does not change in time if we neglect mergers or accretion processes.
Hence, the sky area covered by the caustics on the source plane changes as a consequence of the cosmological expansion only.

The area enclosed by the critical line is
\begin{equation}
A_\text{c} \, = \pi \, \theta_\text{t}^2 \, .    
\end{equation}
This can be interpreted as the cross section for a strong lensing event. 

The average surface mass density within the Einstein ring equals the critical surface density,
\begin{equation}
<\Sigma>_{\theta_\text{t}}  = \Sigma_\text{cr} \equiv \frac{c^2}{4\pi G} \frac{D_\text{s}}{D_\text{l}D_\text{ls}} \, .
\end{equation}
Due to expansion,
\begin{equation}
\frac{d \ln \Sigma_\text{cr}  }{d t_0}= -  \frac{H(z_\text{l})}{1+z_\text{l}} \, .
\end{equation}

Considering that $\theta_t = R_\text{t}/D_\text{l}$, and
\begin{equation}
\frac{d R_\text{t}}{d t_0} =  \frac{d R_\text{t}} {d <\Sigma>_{\theta_\text{t}} } \frac{d <\Sigma>_{\theta_\text{t}} }{d t_0} ,
\end{equation}
the time variation of the angular size of the critical line can be written as
\begin{equation}
\frac{d \theta_\text{t}} {d t_0}=  -\left( \frac{d \theta_\text{t}} {d <\Sigma>_{\theta_\text{t}} } + \theta_\text{t} \right)   \frac{H(z_\text{l})}{1+z_\text{l}} .
\end{equation}

The drift of the cross section can be written as
\begin{equation}
\frac{d \ln A_\text{c}} {d t_0} =  2\frac{d \ln \theta_\text{t}} {d t_0} .    
\end{equation}

Let us consider the axially symmetric lens to be a power-law density (PL) profile,
\begin{equation}
\rho_\text{PL} = \rho_\text{s} \left( \frac{r_\text{s}}{r} \right)^{\gamma_\text{PL}} .
\end{equation}
For a slope $\gamma_\text{PL}=2$, the PL lens reduces to a singular isothermal sphere (SIS).

The critical line of a PL lens lies at
\begin{equation}
\theta_\text{t} =  \left[ 4 \pi    \frac{ \Gamma[ (\gamma_\text{PL}-1)/2] }{\sqrt{\pi} \Gamma[\gamma_\text{PL}/2]} \left( \frac{V_\text{s}}{c} \right)^2 \frac{D_\text{ls}}{D_\text{s}}   \right]^{1/(\gamma_\text{PL}-1)}  \left( \frac{r_\text{s}}{D_\text{l}} \right)^\frac{\gamma_\text{PL}-2}{\gamma_\text{PL}-1},
\end{equation}
where $V_\text{s}$ is the circular velocity at radius $r_\text{s}$.
For a SIS, $\gamma_\text{PL}=2$,
\begin{equation}
\theta_\text{t} = \frac{4 \pi \sigma_\text{v}^2}{c^2} \, \frac{D_\text{ls}}{D_\text{s}} \, ,
\label{eq:theta_E}
\end{equation}
where $\sigma_\text{v}(=V_\text{s})$ is velocity dispersion of the galaxy or galaxy cluster acting as lens.

The variation of the area inside the critical curve after an observation time $\delta t_0$ is 
\begin{equation}
\frac{\delta A_\text{c}}{A_\text{c}} \,  =  \frac{\text{d} \ln A_\text{c}}{\text{d} t} \,  \delta t_0 = - 2 \frac{2- \gamma_\text{PL}}{1- \gamma_\text{PL}} \frac{H(z_\text{l})}{1+z_\text{l}} \, .
\label{eq:Area}
\end{equation}

%


Lenses are approximately isothermal ($\gamma_\text{PL} \sim 2$) 
in the regions where multiple images form and we do not expect significant changes of the cross section with time. However, the slope of the density profile could be shallower in the very inner regions ($\gamma_\text{PL} \lesssim 2$), 
as described by the cuspy dark halo models predicted by numerical simulations \citep{nav+al97}. 
For $1<\gamma_\text{PL}<2$, we expect that high magnification lensing regions get larger in time.


In this scenario, $\delta A_\text{c} > 0$ ($<0$), and sources behind the lens enter (exit) the caustics for shallow (steep) profiles, $1<\gamma_\text{PL} < 2$ ($\gamma_\text{PL} >2$), generating the occurrence (disappearance) of pairs of bright images located close to the critical lines. 


The number $\delta N_\text{SL}$ of new events (that is, appearance of new pairs of images) due to the a change in strong lensing cross section is 
\begin{equation}
\delta N_\text{SL} = \, N_\text{SL} \, \frac{\delta A_\text{c}}{A_\text{c}} \, ,
\label{eq:delta_N} 
\end{equation}
where $N_\text{SL}$ is the number of lenses already present in the survey area. 

Multiply lensed images are routinely observed behind massive galaxies or galaxy clusters, \citep[e.g.,][]{Covone_2006A&A,2015MNRAS.446..683D}. For a next-generation, Euclid-like survey \citep{eucl_lau_11}, we expect $N_\text{SL} \sim 10^5$ strong gravitational lenses. Let us consider a population of self-similar lenses. Dark matter halos are well modelled as Navarro-Frenk-White profiles \citep{nav+al97}, with an inner, shallow, power slope of $\gamma_\text{PL}\sim 1$, and an outer, steep slope of $\gamma_\text{PL}\sim 3$. Considering an uniform slope of $\gamma_\text{PL}\sim 1.5$ in the strong lensing region, for a self-similar population of $10^5$ lenses at $z_\text{l} \sim 0.3$, the probability that a new strong lens appears due to drift in a 10 year observation span is a tiny $\sim 0.01$ per cent.
The presence of a massive central galaxy or baryonic processes can make the inner slope steeper, which would make the lensing cross-section for lensing drift smaller, see Fig.~\ref{fig_delta_NSL_gamma_PL}.

A similar appearance of images might occur even more frequently if taking the granular composition of the lens (such as stars, MACHOs, or dark matter substructures) into account \citep{men+al20}, which turns the smooth caustic into a network of microcaustics. This could increase the cross-section.

In the fortunate event of the formation of a new lensing system by the cosmic drift, we might wait some time to actually see the pair of newly formed images because of the time delay between them. The second image takes more time to reach the observer than the first one. However, the delay is expected to be small since the source has just entered the caustic and the new images form near the critical lines.

Finally, we did not consider here the effects due to the finite size of the source. Indeed, larger sources will require a finite time to cross the slowly expanding caustics.


\subsection{Image drift}

Cosmic parallax is the analogue of redshift drift, but in the transverse direction. It was initially proposed as a probe for anisotropic universes, where the angular separation between distant sources (e.g. quasars) changes in time due to background anisotropic expansion \citep{que+al12}.

In an isotropic expanding universe, transverse motion is generated in lensing systems. The image separation changes in time due to cosmic expansion. 

Angular separations $\Delta \theta$ are of the order of two times the Einstein ring, i.e. $\sim3$ arcseconds for massive galaxies ($\sigma_\text{v}\sim 300~\text{km}~\text{s}^{-1}$), or $\sim30$ arcseconds for clusters of galaxies ($\sigma_\text{v}\sim 1000~\text{km}~\text{s}^{-1}$).

The temporal variation of $\Delta \theta$ can be written as
\begin{equation}
\label{eq_ima_dri_1}
\frac{d \Delta \theta}{dt_0}  \sim 2 \frac{d \theta_\text{t}} {d t_0} \, .
\end{equation}
For a PL profile, 
\begin{equation}
\label{eq_ima_dri_2}
\frac{ \text{d} \ln \theta_\text{t}} {d t_0}= - \left( \frac{\gamma_\text{PL}-2}{\gamma_\text{PL}-1} \right) \frac{H(z_\text{l})}{1+z_\text{l}},
\end{equation}

In $\sim 10$ years, the image separation for a single source behind a lens at $z_\text{l}=0.3$ with $\gamma_\text{PL}\sim 1.5$ and $\sigma_\text{v}\sim 300~(1000)~\mathrm{km/s}$ decreases by $\delta \Delta \theta \sim0.004~(0.04)~\mu\text{as}$. Considering that planned surveys are going to find $\sim 10^5$ strong lenses, an astrometric accuracy of $\sim 1$-$10~\mu\text{as}$ is required to detect the image drift from a sample of self-similar lenses.

Such tiny variations could be within reach of future missions. Operating experiments are already very successful. In its second data release, Gaia has detected nearly 30000 bright quasars ($G < 18~\text{mag}$), with an astrometric accuracy of $\sim100~\mu\text{as}$ \citep{gaia_dr2_mignard+18}.

Very Long Baseline Interferometry (VLBI) astronomy can achieve astrometric accuracy of the order $\sim 100 \mu\text{as}$  at low radio frequency in a single observation. The RadioAstron Space VLBI mission produced images of the quasar BL Lac to an angular resolution of $\sim20\mu\text{as}$ \citep{gom+al16}. The $10 \mu\text{as}$ goal of astrometry at $\sim 1.6$~GHz using VLBI with the Square Kilometer Array (SKA) is expected to be met within this decade \citep{rio+al17}.


\subsection{Time delay drift}

Due to gravitational lensing, multiple images take different times to reach the observer. For a detailed discussion of the change over time of the time delay for general lensing models, we refer to \citet{zi+ei18,wuc+al21}. Here we briefly discuss the drift in the context of future surveys.

The time delay between the two images by a SIS can be written as \citep{ogu+al02}
\begin{equation}
\label{eq_tim_del_1}
c \, \Delta t \, = \, 8\pi \frac{\sigma_\text{v}^2}{c^2} (1+z_\text{l}) D_\text{l} \, \beta  \, .
\end{equation}
Due to cosmic expansion, the time delay changes. The time variation is
\begin{equation}
\label{eq_tim_del_2}
\frac{d \ln \Delta t}{d t_0}= H_0 \, .
\end{equation}
There is no apparent dependence on the source redshift  in Eq.~(\ref{eq_tim_del_2}), but the condition $\beta <\theta_\text{t}$, where $\theta_\text{t}$ depends on redshifts, has still to be fulfilled in order to have a multiple image system.

The relative variation in the time-delay over a time span $\delta t_0$ is
\begin{equation}
\label{eq_tim_del_3}
\frac{\delta\Delta t}{\Delta t}= H_0 \delta t_0 \, .
\end{equation}
Due to cosmic expansion, the variation is of order of $\sim10^{-9}$ in 10 years. Time delays in excess of one year have already been recorded, which implies a time-delay drift of $\sim 0.02~\text{s}$ in ten years.

Time-delays have been measured with a percent level accuracy \citep{tr+ma16}. Even considering the monitoring of $\sim 10^5$ lensing systems, as reachable in future wide surveys, present day accuracy has to be improved by $\sim4$-$5$ orders of magnitude in order to detect the effect of the redshift drift on the time delay in the sample.

It has been shown that change in time delay over time between pairs of multiply imaged events of lensed, repeating FRBs offers a more convenient observable for real time cosmology \citep{zi+ei18,wuc+al21}.

\subsection{The drifted shear}

The  distortion of the background source due to lensing can be quantified in terms of the tangential shear. For a PL,
\begin{equation}
\gamma_+ = \frac{\gamma_\text{PL}-1}{2} \left( \frac{\theta_\text{t}}{\theta} \right)^{\gamma_\text{PL}-1} .
\end{equation}
Considering Eq.~(\ref{eq_ima_dri_2}), the drift of the shear can be written as
\begin{equation}
\frac{ \text{d} \ln \gamma_+} { \text{d} t_0} =
(2 - \gamma_\text{PL})  \frac{H(z_\text{l})}{1+z_\text{l}},
\end{equation}
At a given angular position $\theta$, the measured shear is constant in time for an isothermal profile ($\gamma_\text{PL}=2$). The lens profile is expected to be steeper than isothermal ($\gamma_\text{PL}\lesssim 3$) in the outer regions where the shear is usually detected over large samples of background galaxies. For such lenses, the shear is expected to get smaller with time. 
The relative shear variation in a ten years span is of the order of $\delta  \gamma_+/\gamma_+ \sim - 10^{9}$. 

The  shear measurement is very challenging. Massive gravitational lenses (as galaxy clusters) generate a tangential shear of $\gamma_+ \sim 10^{-3}$ at the intermediate radii where the signal can be measured \citep{ume+al14,wtg_III_14,ser+al18_psz2lens}.

Detection of time variations seems beyond the reach of next generation surveys. The shear accuracy of a survey can be approximated as
\begin{equation}
    \sigma_{\gamma} \sim \frac{\sigma_\text{e}}{\sqrt{N_\text{bkg}}} \, ,
\end{equation}
where $\sigma_e (\sim 0.4)$  is the dispersion in intrinsic ellipticity of background sources and $N_\text{bkg}$ is the number of galaxies with measured shear. A next-generation, wide survey can scan half of the sky and detect $\sim 30$ galaxies per squared arcminute. This implies an exquisite accuracy of $\sigma_{\gamma} \sim 10^{-5}$, still not enough to detect the effect of the cosmological drift on the shear.


\section{Competing effects}

The case of comoving observer, lens, and sources, and lens with static mass distribution is ideal. Transverse proper motion, mass accretion, growth of density perturbations can modify a lens system over time \citep{zi+ei18,wuc+al21}. 

Competing effects should be disentangled to observe cosmic expansion. Let us briefly consider the size of some effects. For a detailed discussion we refer to \citet{zi+ei18,wuc+al21}.

A source can enter or exit the caustics of a stationary lens due to its peculiar motion,
\begin{equation}
\frac{ \delta A_\text{c}}{A_\text{c}} = \frac{2 v_\beta \, \delta t_0}{\theta_\text{t}},
\end{equation}
where $v_\beta$ is the transverse angular velocity of the source. To observe a new strong lensing event in a next-generation campaign in a ten years span, all sources should coherently move with a transverse velocity of $\sim10^4~\text{km/s}$. The transverse motion effect can be larger than the variation due to the radial drift but there are ways to disentangle the two phenomena \citep{zi+ei18}, e.g. by considering the the different pairs of time delays in quadruple systems \citep{wuc+al21}. 
The impact of transverse motion on strongly lensed system, even though not in the context of cosmological effects, was also discussed in \citet{da+lu17}.

Transverse motion would also play a relevant role for microimages produced by the granular distribution of the gravitational lens. 

The mass distribution of a lens can change over time due to mass accretion from the environment, mergers with nearby halos, or baryonic processes in the inner regions. The formation of a halo can be seen as as a two-stage process \citep{di+jo19}. In the early fast-accretion regime, the halo grows rapidly and its profile maintains a roughly universal shape. In a second phase, the growth slows down, accretion only affects the outer regions, and the scale radius of the halo approaches a constant value. As a result, the inner regions of the halo remains more or less static in physical coordinates. If mass is accreted in the inner regions, the cross section for lensing changes (for the SIS) as
\begin{equation}
\frac{\delta A_\text{c}}{A_\text{c}} = \frac{2\delta m_\text{cyl,t}}{m_\text{cyl,t}},
\end{equation}
where $m_\text{cyl,t}$ is the projected mass within the Einstein radius. To observe a new strong lensing event in a next-generation campaign in a ten year span, sources should coherently change their mass by a substantial $\sim5\times 10^{-4}$ per cent.

The discussed effects are quite small but they could be larger than the drift effect. However, they could be differentiated due to their random nature. Sources are not expected to move in the same direction out of the Hubble flow. For lenses with a self-similar structure, the cosmic expansion coherently modify the caustics whereas peculiar motions could either reduce or enlarge them. The mass in the inner region could both increase by mergers or accretion or decrease due to baryonic processes such feedback by the active galactic nucleus that could push the gas in the outer regions.


\section{Conclusions}

In recent years, real-time cosmology has been proposed as a viable even though challenging novel research domain \citep{que+al12}. Improved astrometric and spectroscopic techniques could make feasible the measurement of the temporal change of radial and transverse position of sources in the sky over relatively short time intervals

New lensed images could pop in the sky due to redshift drift. This is a peculiar signature of the cosmological expansion. Differently from lensing of transient events such as supernovae \citep{kel+al15}, newly formed images due to drift would not disappear.

New lensing images are due to the caustic area being made larger by the drift. The larger caustic can cover a previously not strongly lensed source which can then be multiply imaged. 
This phenomenon could be characterized by a non negligible transition time in which the source would be crossing the caustic and thus (if small enough) extremely magnified. This is of particular interest given the recent detections of highly magnified stars in strong lensing events \citep{kel+al15, Kaurov2019ApJ, Welch_2022}.

The caustic shape could change also for internal motions of the lens constituents, as can happen in minor or major mergers, or a source in proper motion could enter the caustic. These events could generate multiple images too and could be difficult to distinguish from lensing drift. However, signatures by cosmic drift or internal and proper motions differ in one major facet. Whereas the cosmic drift acts coherently and in the same direction on the full lens population, peculiar motions act randomly. Cosmic drift could be difficult to distinguish from peculiar motions on a case by case basis, but its signature could be recovered in a statistical lensing sample.

The lensing system acts as a converter. The cosmic expansion modifies the geometry of the system. The redshift drift of the lens and the source are transformed into a variation of angular separation or time intervals of the images. This could open new avenues to detection.

Our proof-of-concept study shows that there are unique lensing phenomena associated to cosmic drift. We considered comoving observers, lens, and sources with static, smooth mass distributions and no proper motions. Systematics and noise in a not idealized setting can complicate the analysis and the detection.

The probability to observe lensing drift events is very tiny. However, the change in time of FRBs has been already emerging as a feasible observational target \citep{zi+ei18,wuc+al21}. 
%
%
As noted by \cite{zi+ei18}, the duration of the burst is a smaller fraction of the time delay between multiple images of a source gravitationally lensed by a galaxy than the human lifetime is to the age of the universe. 
Thus, repeating, strongly lensed FRBs may offer a target for observing cosmological evolution in real time.

Here, we discussed some novel lensing phenomena. Many newly predicted lensing features have an historic tradition to appear unfeasible from the experimental point of view when first discussed. Albert Einstein himself doubted the possibility to directly observe the deviation of light by nearby stars other than the Sun \citep{ein36}, which is now detected \citep{sah+al17}. Observations have proved pessimists wrong a number of times. 
Next generation surveys with $\mu$arcsec angular resolution will show whether this might be the case for the lensing drift phenomena.

\section*{Acknowledgements}
We thank Adi Zitrin for insightful discussions and for pointing out the role of microlensing.
We thank Valerio Busillo for his comments and drafting of the Figure 1.
MS acknowledges financial contribution from contracts ASI-INAF n.2017-14-H.0 and INAF mainstream project 1.05.01.86.10.

\section*{Data Availability}
No new data were generated or analysed in support of this research. All data used in this paper were obtained from the cited literature.



\bibliographystyle{mnras}
\bibliography{biblio} 

\providecommand{\noopsort}[1]{}\providecommand{\singleletter}[1]{#1}%
\begin{thebibliography}{}
\makeatletter
\relax
\def\mn@urlcharsother{\let\do\@makeother \do\$\do\&\do\#\do\^\do\_\do\%\do\~}
\def\mn@doi{\begingroup\mn@urlcharsother \@ifnextchar [ {\mn@doi@}
  {\mn@doi@[]}}
\def\mn@doi@[#1]#2{\def\@tempa{#1}\ifx\@tempa\@empty \href
  {http://dx.doi.org/#2} {doi:#2}\else \href {http://dx.doi.org/#2} {#1}\fi
  \endgroup}
\def\mn@eprint#1#2{\mn@eprint@#1:#2::\@nil}
\def\mn@eprint@arXiv#1{\href {http://arxiv.org/abs/#1} {{\tt arXiv:#1}}}
\def\mn@eprint@dblp#1{\href {http://dblp.uni-trier.de/rec/bibtex/#1.xml}
  {dblp:#1}}
\def\mn@eprint@#1:#2:#3:#4\@nil{\def\@tempa {#1}\def\@tempb {#2}\def\@tempc
  {#3}\ifx \@tempc \@empty \let \@tempc \@tempb \let \@tempb \@tempa \fi \ifx
  \@tempb \@empty \def\@tempb {arXiv}\fi \@ifundefined
  {mn@eprint@\@tempb}{\@tempb:\@tempc}{\expandafter \expandafter \csname
  mn@eprint@\@tempb\endcsname \expandafter{\@tempc}}}

\bibitem[\protect\citeauthoryear{{Applegate} et~al.,}{{Applegate}
  et~al.}{2014}]{wtg_III_14}
{Applegate} D.~E.,  et~al., 2014, \mn@doi [\mnras] {10.1093/mnras/stt2129},
  \href {http://adsabs.harvard.edu/abs/2014MNRAS.439...48A} {439, 48}

\bibitem[\protect\citeauthoryear{{Covone}, {Kneib}, {Soucail}, {Richard},
  {Jullo}  \& {Ebeling}}{{Covone} et~al.}{2006}]{Covone_2006A&A}
{Covone} G.,  {Kneib} J.~P.,  {Soucail} G.,  {Richard} J.,  {Jullo} E.,
  {Ebeling} H.,  2006, \mn@doi [\aap] {10.1051/0004-6361:20053384}, \href
  {https://ui.adsabs.harvard.edu/abs/2006A&A...456..409C} {456, 409}

\bibitem[\protect\citeauthoryear{{Dai} \& {Lu}}{{Dai} \& {Lu}}{2017}]{da+lu17}
{Dai} L.,  {Lu} W.,  2017, \mn@doi [\apj] {10.3847/1538-4357/aa8873}, \href
  {https://ui.adsabs.harvard.edu/abs/2017ApJ...847...19D} {847, 19}

\bibitem[\protect\citeauthoryear{{Diego} et~al.,}{{Diego}
  et~al.}{2015}]{2015MNRAS.446..683D}
{Diego} J.~M.,  et~al., 2015, \mn@doi [\mnras] {10.1093/mnras/stu2064}, \href
  {https://ui.adsabs.harvard.edu/abs/2015MNRAS.446..683D} {446, 683}

\bibitem[\protect\citeauthoryear{{Diemer} \& {Joyce}}{{Diemer} \&
  {Joyce}}{2019}]{di+jo19}
{Diemer} B.,  {Joyce} M.,  2019, \mn@doi [\apj] {10.3847/1538-4357/aafad6},
  \href {https://ui.adsabs.harvard.edu/abs/2019ApJ...871..168D} {871, 168}

\bibitem[\protect\citeauthoryear{{Einstein}}{{Einstein}}{1936}]{ein36}
{Einstein} A.,  1936, \mn@doi [Science] {10.1126/science.84.2188.506}, \href
  {https://ui.adsabs.harvard.edu/abs/1936Sci....84..506E} {84, 506}

\bibitem[\protect\citeauthoryear{{Gaia Collaboration} et~al.,}{{Gaia
  Collaboration} et~al.}{2018}]{gaia_dr2_mignard+18}
{Gaia Collaboration} et~al., 2018, \mn@doi [\aap]
  {10.1051/0004-6361/201832916}, \href
  {https://ui.adsabs.harvard.edu/abs/2018A&A...616A..14G} {616, A14}

\bibitem[\protect\citeauthoryear{{G{\'o}mez} et~al.,}{{G{\'o}mez}
  et~al.}{2016}]{gom+al16}
{G{\'o}mez} J.~L.,  et~al., 2016, \mn@doi [\apj] {10.3847/0004-637X/817/2/96},
  \href {https://ui.adsabs.harvard.edu/abs/2016ApJ...817...96G} {817, 96}

\bibitem[\protect\citeauthoryear{{Harrison}}{{Harrison}}{2000}]{har00}
{Harrison} E.~R.,  2000, {Cosmology. The science of the universe.}.
Cambridge University Press

\bibitem[\protect\citeauthoryear{{Kaurov}, {Dai}, {Venumadhav},
  {Miralda-Escud{\'e}}  \& {Frye}}{{Kaurov} et~al.}{2019}]{Kaurov2019ApJ}
{Kaurov} A.~A.,  {Dai} L.,  {Venumadhav} T.,  {Miralda-Escud{\'e}} J.,   {Frye}
  B.,  2019, \mn@doi [\apj] {10.3847/1538-4357/ab2888}, \href
  {https://ui.adsabs.harvard.edu/abs/2019ApJ...880...58K} {880, 58}

\bibitem[\protect\citeauthoryear{{Kelly} et~al.,}{{Kelly}
  et~al.}{2015}]{kel+al15}
{Kelly} P.~L.,  et~al., 2015, \mn@doi [Science] {10.1126/science.aaa3350},
  \href {https://ui.adsabs.harvard.edu/abs/2015Sci...347.1123K} {347, 1123}

\bibitem[\protect\citeauthoryear{{Laureijs} et~al.,}{{Laureijs}
  et~al.}{2011}]{eucl_lau_11}
{Laureijs} R.,  et~al., 2011, arXiv:1110.3193, \href
  {http://adsabs.harvard.edu/abs/2011arXiv1110.3193L} {}

\bibitem[\protect\citeauthoryear{{Liu}, {Li}, {Gao}  \& {Zhu}}{{Liu}
  et~al.}{2019}]{liu+al19}
{Liu} B.,  {Li} Z.,  {Gao} H.,   {Zhu} Z.-H.,  2019, \mn@doi [\prd]
  {10.1103/PhysRevD.99.123517}, \href
  {https://ui.adsabs.harvard.edu/abs/2019PhRvD..99l3517L} {99, 123517}

\bibitem[\protect\citeauthoryear{{Loeb}}{{Loeb}}{1998}]{loe98}
{Loeb} A.,  1998, \mn@doi [\apjl] {10.1086/311375}, \href
  {https://ui.adsabs.harvard.edu/abs/1998ApJ...499L.111L} {499, L111}

\bibitem[\protect\citeauthoryear{{Meneghetti} et~al.,}{{Meneghetti}
  et~al.}{2020}]{men+al20}
{Meneghetti} M.,  et~al., 2020, \mn@doi [Science] {10.1126/science.aax5164},
  \href {https://ui.adsabs.harvard.edu/abs/2020Sci...369.1347M} {369, 1347}

\bibitem[\protect\citeauthoryear{{Navarro}, {Frenk}  \& {White}}{{Navarro}
  et~al.}{1997}]{nav+al97}
{Navarro} J.~F.,  {Frenk} C.~S.,   {White} S.~D.~M.,  1997, \mn@doi [\apj]
  {10.1086/304888}, \href {http://ads.astro.puc.cl/abs/1997ApJ...490..493N}
  {490, 493}

\bibitem[\protect\citeauthoryear{{Oguri}, {Taruya}, {Suto}  \&
  {Turner}}{{Oguri} et~al.}{2002}]{ogu+al02}
{Oguri} M.,  {Taruya} A.,  {Suto} Y.,   {Turner} E.~L.,  2002, \mn@doi [\apj]
  {10.1086/339064}, \href
  {https://ui.adsabs.harvard.edu/abs/2002ApJ...568..488O} {568, 488}

\bibitem[\protect\citeauthoryear{{Pasquini} et~al.,}{{Pasquini}
  et~al.}{2005}]{pas+al05}
{Pasquini} L.,  et~al., 2005, The Messenger, \href
  {https://ui.adsabs.harvard.edu/abs/2005Msngr.122...10P} {122, 10}

\bibitem[\protect\citeauthoryear{{Piattella} \& {Giani}}{{Piattella} \&
  {Giani}}{2017}]{pi+gi17}
{Piattella} O.~F.,  {Giani} L.,  2017, \mn@doi [\prd]
  {10.1103/PhysRevD.95.101301}, \href
  {https://ui.adsabs.harvard.edu/abs/2017PhRvD..95j1301P} {95, 101301}

\bibitem[\protect\citeauthoryear{{Quercellini}, {Amendola}, {Balbi}, {Cabella}
  \& {Quartin}}{{Quercellini} et~al.}{2012}]{que+al12}
{Quercellini} C.,  {Amendola} L.,  {Balbi} A.,  {Cabella} P.,   {Quartin} M.,
  2012, \mn@doi [\physrep] {10.1016/j.physrep.2012.09.002}, \href
  {https://ui.adsabs.harvard.edu/abs/2012PhR...521...95Q} {521, 95}

\bibitem[\protect\citeauthoryear{{Rioja}, {Dodson}, {Orosz}, {Imai}  \&
  {Frey}}{{Rioja} et~al.}{2017}]{rio+al17}
{Rioja} M.~J.,  {Dodson} R.,  {Orosz} G.,  {Imai} H.,   {Frey} S.,  2017,
  \mn@doi [\aj] {10.3847/1538-3881/153/3/105}, \href
  {https://ui.adsabs.harvard.edu/abs/2017AJ....153..105R} {153, 105}

\bibitem[\protect\citeauthoryear{{Sahu} et~al.,}{{Sahu}
  et~al.}{2017}]{sah+al17}
{Sahu} K.~C.,  et~al., 2017, \mn@doi [Science] {10.1126/science.aal2879}, \href
  {https://ui.adsabs.harvard.edu/abs/2017Sci...356.1046S} {356, 1046}

\bibitem[\protect\citeauthoryear{{Sandage}}{{Sandage}}{1962}]{san62}
{Sandage} A.,  1962, \mn@doi [\apj] {10.1086/147385}, \href
  {https://ui.adsabs.harvard.edu/abs/1962ApJ...136..319S} {136, 319}

\bibitem[\protect\citeauthoryear{{Schneider}, {Ehlers}  \& {Falco}}{{Schneider}
  et~al.}{1992}]{sef92}
{Schneider} P.,  {Ehlers} J.,   {Falco} E.~E.,  1992, {Gravitational Lenses}.
{Springer-Verlag Berlin Heidelberg}, \mn@doi{10.1007/978-3-662-03758-4}

\bibitem[\protect\citeauthoryear{{Sereno} et~al.,}{{Sereno}
  et~al.}{2018}]{ser+al18_psz2lens}
{Sereno} M.,  et~al., 2018, \mn@doi [Nature Astronomy]
  {10.1038/s41550-018-0508-y}, \href
  {https://ui.adsabs.harvard.edu/abs/2018NatAs...2..744S} {2, 744}

\bibitem[\protect\citeauthoryear{{Treu} \& {Marshall}}{{Treu} \&
  {Marshall}}{2016}]{tr+ma16}
{Treu} T.,  {Marshall} P.~J.,  2016, \mn@doi [\aapr]
  {10.1007/s00159-016-0096-8}, \href
  {https://ui.adsabs.harvard.edu/abs/2016A&ARv..24...11T} {24, 11}

\bibitem[\protect\citeauthoryear{{Umetsu} et~al.,}{{Umetsu}
  et~al.}{2014}]{ume+al14}
{Umetsu} K.,  et~al., 2014, \mn@doi [\apj] {10.1088/0004-637X/795/2/163}, \href
  {http://adsabs.harvard.edu/abs/2014ApJ...795..163U} {795, 163}

\bibitem[\protect\citeauthoryear{Welch et~al.,}{Welch
  et~al.}{2022}]{Welch_2022}
Welch B.,  et~al., 2022, \mn@doi [Nature] {10.1038/s41586-022-04449-y}, 603,
  815

\bibitem[\protect\citeauthoryear{{Wucknitz}, {Spitler}  \& {Pen}}{{Wucknitz}
  et~al.}{2021}]{wuc+al21}
{Wucknitz} O.,  {Spitler} L.~G.,   {Pen} U.~L.,  2021, \mn@doi [\aap]
  {10.1051/0004-6361/202038248}, \href
  {https://ui.adsabs.harvard.edu/abs/2021A&A...645A..44W} {645, A44}

\bibitem[\protect\citeauthoryear{{Zitrin} \& {Eichler}}{{Zitrin} \&
  {Eichler}}{2018}]{zi+ei18}
{Zitrin} A.,  {Eichler} D.,  2018, \mn@doi [\apj] {10.3847/1538-4357/aad6a2},
  \href {https://ui.adsabs.harvard.edu/abs/2018ApJ...866..101Z} {866, 101}

\makeatother
\end{thebibliography}








\bsp	
\label{lastpage}
\end{document}